\documentclass[prd,preprintnumbers,twocolumn,amsmath,amssymb,nofootinbib,showpacs]{revtex4}
\def\scn#1#2{\section{#1}\lb{#2}}
\def\sscn#1#2{\subsection{#1}\lb{#2}}

\usepackage{epsfig,amssymb,amsfonts,verbatim,enumerate}

\def\cmo{\mu}

\def\bfl{\begin{flushleft}}
\def\efl{\end{flushleft}}
\def\bfr{\begin{flushright}}
\def\efr{\end{flushright}}
\def\bc{\begin{center}}
\def\ec{\end{center}}
\def\be{\begin{equation}}
\def\ee{\end{equation}}
\def\ba{\begin{eqnarray}}
\def\ea{\end{eqnarray}}
\def\baa#1{\begin{array}{#1}}
\def\eaa{\end{array}}
\def\bw{\begin{widetext}}
\def\ew{\end{widetext}}

\def\lb#1{\label{#1}}
\def\bit{\begin{itemize}} 
\def\eit{\end{itemize}}

\def\schrod{Schr\"odinger  }

\def\lan{{\cal L}}

\def\Sign#1{\, \text{sign}\left(#1\right) }

\newtheorem{Def}{Def.}[section]

\newtheorem{Prp}[Def]{Proposition}

\newcommand{\IR}{\text{\rm I \hspace{-.8 em} R}}

\begin{document}

\preprint{Acta Phys. Polon. B 42 (2011) 261-292 [arXiv: 0912.4139]}

\title{
Spontaneous symmetry breaking and mass generation 
as built-in phenomena 
in logarithmic nonlinear quantum theory
}

\author{Konstantin G. Zloshchastiev}

\affiliation{
Department of Physics and Center for Theoretical Physics,
University of the Witwatersrand,
Johannesburg, 2050, South Africa}



\date{21 Dec 2009 (ArXiv), 29 Sep 2010 (APPB)}


\begin{abstract}
Our primary task  is to
demonstrate  that the logarithmic nonlinearity
in the quantum wave equation
can cause the spontaneous symmetry breaking 
and mass generation phenomena on its own,
at least in principle. 
To achieve this goal, 
we view the physical vacuum as a kind of the fundamental
Bose-Einstein condensate  embedded
into the fictitious Euclidean space.
The
relation of such description 
to that of the physical (relativistic) observer 
is established via the fluid/gravity correspondence map, the
related issues, such as the induced gravity and scalar field, relativistic
postulates, Mach's principle and cosmology,
are discussed.
For estimate the values of the generated masses of the otherwise
massless particles such as the photon,
we propose few simple models 
which take into account  small vacuum
fluctuations.
It turns out that 
the photon's mass can be naturally expressed
in terms of the elementary electrical charge
and the extensive length parameter 
of the nonlinearity.
Finally, we outline 
the topological properties 
of the logarithmic theory and
corresponding solitonic solutions.
\end{abstract}



\maketitle



\scn{Introduction}{sec-i}

Current observational data in astrophysics 
are probing a regime of departures from classical relativity with sensitivities that are relevant for the study of the quantum-gravity problem
\cite{Collaborations:2009zq,Stecker:2009hj}.
On the other hand, the quantum theory of gravity which would be
both widely agreed upon and
capable of making unique testable predictions is still pending.
In this connection, the effective non-axiomatic
theories and semi-phenomenological approaches
guided by the physical intuition
can be very helpful as they may provide new ideas and insights
\cite{AmelinoCamelia:2008qg}.
We already proposed elsewhere
\cite{Zloshchastiev:2009zw}
that the nontrivial vacuum 
causes the deformation of the quantum
wave equations of the universal form: 
\be\lb{e-xmain}
\left[
\hat{{\cal H}} 
-
\beta^{-1} \ln{(\Omega |\Psi|^2)}
\right]
\Psi
= 0,
\ee
where 
$\Psi$ refers in general to the complex-valued wave functional
and
$\hat{{\cal H}} $ is the operator which form
is determined by 
a physical setup.
The physical motivation behind this equation as well as its
unique properties are listed in the Appendix.
Here $\beta$ and $\Omega$ are constant parameters.
If we impose that $\Omega$ has the dimensionality of a spatial volume
then the logarithmic term (\ref{e-xmain})
introduces the primary (extensive) length scale,
\be
\ell_\Omega = \Omega^{1/(D-1)},
\ee
which role and possible physical meaning will be discussed below;
here $D-1$ refers to the number of spatial dimensions, throughout
the paper it is assumed $D =4$. 

It was shown
that 
some phenomenological consequences 
of such theory
are actually model-independent and can be derived even
at the kinematical level, i.e., prior to specifying the dynamical
details of a quantum-gravitational model. 
One of the primary phenomenological implications of this theory
is that for any two freely-moving particles
the following relation is valid 
\be\lb{e-b12}
\frac{d\tau_2}{d\tau_1}
=
\frac{E_2 - E_0}{E_1 - E_0}
=
1
-
\frac{\Delta E}{E_0}
+
{\cal O} (E^2/E_0^2)
,
\ee
where $\tau_i$ and $E_i$ are the proper time and energy of
the $i$th particle,
$E_0$ is the energy of the vacuum of a theory; for the 
vacuum not affected by external fields
that would be
$E_0 = \pm E_{\text{QG}}$, $E_{\text{QG}} \lesssim 10^{19}$ GeV.
The value $E_0$ 
defines another length scale, the Compton-type one:
\be
\ell_0 = h c/|E_0|,
\ee
which value thus can be as small as the Planck one.
We expect that the properties of any dynamical systems
immersed in such vacuum can change drastically
when their characteristic length scales approach either of the critical
values $\ell_0$ and $\ell_\Omega$, or, in terms of energy,
$E_0$ and $E_\Omega  = h c/\ell_\Omega$. 

The effective refractive index can be directly
computed from corresponding dispersion relations
(taking into account that both the Planck relation and energy additivity
of uncorrelated systems
survive
in the logarithmic theory \cite{BialynickiBirula:1976zp}, in contrast to other nonlinear
extensions of quantum mechanics, see also the Appendix).
In the Cauchy form the index can be written as
\be
n^2 
=
1 + \cmo_\gamma 
\left[1+ {\cal M}(\omega) (\omega/2 \pi c)^2
\right]
,
\ee
where
$
\cmo_\gamma = \chi_\gamma^2 -1 
$
and
${\cal M} (\omega)
=
(2 \pi c/\omega_0)^2
\left(
1
\pm
2 
\omega_0 / \omega
\right)$
are, respectively, the constant of
refraction
and
dispersion coefficient
of the vacuum,
$\omega 
$ is the angular frequency of the electromagnetic wave,
$\omega_0 = |E_0|/\hbar$ is the proper frequency of the vacuum,
$\pm = -\Sign{E_0}$.

All this suggests that 
the vacuum is the medium with non-trivial
properties which affects photons and other particles propagating through it,
and the effects grow along with particles' energies. 
The predicted phenomena which can be derived from Eq. (\ref{e-b12}) 
can be cast into three groups:
\begin{enumerate}[(i)]
\item {\it subluminal phenomena}:
the estimates imply that the particles with higher energy propagate slower than those with lower one, therefore, for a high-energy particle the mean free path, lifetime in a high-energy state and, therefore, travel distance from the source can be significantly larger than one would expect from the conventional relativity theory.
There already exists tentative
evidence of this effect, often referred as the ``high-energy tail'' \cite{Kifune:1999ex};
\item {\it transluminal phenomena}: 
according to the theory,
particles can reach the speed of light in
 vacuum at finite energy.
This may cause 
the ``luminal boom'' in vacuum and
appearance of a conical front of the
Cherenkov-type shock wave. 
These effects can be detected at the Earth's particle accelerators - 
the special feature 
of the latter 
is 
the particles get accelerated to 
ultrarelativistic speeds in a controlled way
whereas the cosmic-ray particles have been accelerated somewhere
else, usually very far from our detectors.
Of course, the outcomes of the accelerator studies will
totally depend on the value of $E_0$
which is not that simple to compute 
because the  vacuum inside the 
accelerator pipe
is distorted
by external fields;
\item {\it superluminal phenomena}:
unlike the tachyons in the classical relativity,
in the logarithmic theory the energies of the superluminal particles 
are real-valued
and stay finite when their propagation speed approaches $c$.
The electromagnetic component
of their Cherenkov radiation may
exhibit the anomalous Doppler effect - similar to the one
for the superluminal (non-point) sources in vacuum 
which was predicted even at the classical relativistic level 
by Bolotovskii and Ginzburg \cite{Bolotovskii:1972ve}. 
Also 
there may exist the phenomenon of
mimicking the double-lobed radio sources 
in astrophysics. 
In general,
the current understanding of physical phenomena
happening in supernovae,
active galactic nuclei and gamma-ray bursts
may need a serious revision.
\end{enumerate}

As mentioned earlier, these phenomena are determined 
mainly
by the kinematics of the theory - 
in a sense, they are analogues 
of the kinematic effects of special relativity.
What about the dynamical ones, is it possible to find any
without specifying an underlying 
microscopical model?
In general the answer 
is naturally ``no'' but there exists (at least)
one exception: the mechanism of the spontaneous symmetry breaking
is actually hidden in the logarithmic term itself.
Of course, this does not 
exclude the existence of other symmetry-breaking
mechanisms caused by the dynamics of a 
concrete 
model.

Spontaneous symmetry breaking occurs when the ground state of a system does not possess the full symmetry of the theory.
The most famous its realization in physics is known
as the (Englert-Brout-)Higgs(-Guralnik-Hagen-Kibble-Nambu-Anderson) mechanism \cite{Englert:1964et}.
The closely related phenomenon is the mass generation 
which has been employed in the Glashow-Weinberg-Salam electroweak theory as to 
explain the nonzero masses of the intermediate vector bosons 
by breaking the electroweak symmetry group $SU(2)\times U(1)$ down to 
the electromagnetic $U(1)$ \cite{Novaes:1999yn}.
This mechanism is mediated by the yet undiscovered particle, Higgs boson,
which mass is currently narrowed to be between $114$ and $158$ GeV -
provided that the Standard Model (SM) remains valid at that energy range.

Despite the overall success of the electroweak theory, 
few questions about its Higgs mechanism remain open.
The one of them is the following.
Intuitively one would expect that anything related to the mass creation
must be governed by gravity, be it classical or quantum -
as the Mach's principle suggests, for instance.
But SM, in its current
formulation, does not have the gravitational sector.
Instead, the role of the ``mass generator'' is transferred to the
Higgs particle from the electroweak sector.
The gravity seems to be totally excluded from this process.
From the mathematical point of view, 
no mass generation mechanism which would naturally
appear as a solely (quantum-)gravitational effect,
i.e., without involving other matter fields, 
has been proposed so far, to our best knowledge.
On the other hand,
in quantum field theory it has been already known 
that the radiation corrections themselves
can 
cause the spontaneous symmetry breaking
\cite{Coleman:1973jx}.

This issue is closely related to the second question - what
{\it is} the physical vacuum: what are its properties,
how do they change at higher energies and shorter scales of length, {\it etc}.
Regrettably, up to now no reliable theory of the physical vacuum actually exists. 
The two most popular nowadays theories, SM and string theory, are practically useless in this regard. 
The former is the operational Lorentz-invariant renormalizable theory which means 
that it does not take into account that the physical vacuum can break the Lorentz invariance at high energies (of order TeV and above) and shorter length scales, also the
theory replaces important parameters, such as masses and charges of elementary particles, by their experimentally measured values thus giving no theoretical explanations for why their values are the way they are. 
In particular, the value of zero-point energy when computed in the electroweak or QCD sectors
disagrees with the one restricted by
astrophysical observations by more than a hundred orders of magnitude -
one of
the most striking manifestations of the so-called ``vacuum catastrophe'' or 
``cosmological constant problem'' noticed by Nernst  
almost a century ago \cite{nernst16}.
The superstring theory, apart from being based on the Lorentz symmetry
too, suffers 
from the so-called ``landscape problem'': 
it gives almost infinitely many mutually exclusive predictions about the structure of the physical vacuum.
It may turn out that this problem is not just a temporary difficulty of the theory 
but 
the indication of 
the Lorentz symmetry's breakdown in Nature at some energy and length scale.
As a result, certain mathematical constructions heavily relying
upon (or motivated by) this symmetry, such as supersymmetry or 
tensor representations of the Poincar\'e group,
should be attributed to the real world with utmost care - 
as their characteristic energy scales can lie outside the validity range 
of the Lorentz-symmetric approach.

The third issue is the mass of the photon.
In the current Standard Model the photon is assumed to be
strangely exceptional - its mass remains zero even after
the electroweak symmetry breaking.
On the other hand,  recent observational data
bring certain evidence that the photon propagates with the subluminal speed
and thus can be assigned a mass, at least effectively, 
but of
an extremely small value, as compared to that of the intermediate vector bosons.
This suggests that the mass generation mechanism for the photon
must be in something drastically different from the electroweak one.

The fourth, last for the moment, issue is almost obvious to guess:
if the electroweak Higgs boson does exist what is the mechanism which
generates its mass?

Thus, regardless of whether the electroweak Higgs particle exists or not,
there should be at least one mass generation mechanism which
lies outside the scope of the Glashow-Weinberg-Salam theory.
What about the logarithmic nonlinearity, can it help
in understanding these problems?
Also, once we have established that the particles freely
propagating in the logarithmic theory
can be effectively viewed as propagating in some non-trivial background
medium,
what is the physical nature of this medium?

\scn{Spontaneous symmetry breaking 
}{sec-ssb}

The first thing to notice is if
in some representation the operator $\hat{{\cal H}} $
can be written as a second-order differential operator
with respect to some variable $X$, i.e.,
$
\hat{{\cal H}} \sim f_1 \frac{\partial^2 ~}{\partial X^2} 
+
f_2 \frac{\partial ~}{\partial X}
$
(we assume $f_1 > 0$ otherwise one must invert the sign of $\beta$
or perform the Wick rotation of $X$)
then
the wave equation (\ref{e-xmain}) can be viewed as 
the equation
of motion of the fictitious particle 
moving on a plane 
$\{\Re (\Psi), \, \Im (\Psi) \}$
in the rotationally-invariant external 
potential 
\be\lb{e-pot}
{ V} (\Psi)
=
\beta^{-1} 
\biggl\{
\Omega |\Psi|^2
\left[
\ln{(\Omega |\Psi|^2)}
-1
\right]
+ 1
\biggr\}
+ 
{ V}_0
,
\ee
where ${ V}_0 \equiv {V} (|\Psi| =1/\sqrt{\Omega})$,
with the role of time coordinate being assigned to $X$ or to $i X$,
as in the semi-classical approach.
It is not difficult to check that
for positive $\beta$ and $\Omega$
this potential has the Mexican-hat shape:
its local maximum is located at 
$|\Psi|=0$ whereas the degenerate minima lie
on the circle $|\Psi|= 
1/\sqrt{\Omega}$
where
the energy of the ``particle''  reaches its minimum.

To present things in a more rigorous way
we use the ideology of the
Bogoliubov-Ginzburg-Landau(-Gross-Pitaevskii) mean-field approach 
\cite{Bogoliubov47}
which is  a special case of the \schrod field method
and originates from the following idea.
Suppose $\Psi$ is originally the functional on a
space of field operators $\hat\psi_{(i)}$ which maps this space onto 
the field of c-numbers. 
As long as those fields themselves depend
on space and time variables $x$
then in certain cases, for instance, when they describe 
identical particles in the same state,
the functional
$\Psi [\hat\psi_{(i)}(x)]$ can be replaced by the function $\Psi (x)$.
The latter is nothing but the probability amplitude which complex square
is a measurable quantity but now the wave equation it satisfies
is not necessarily linear.
This $\Psi (x)$ is traditionally called the wave function of the 
Bose-Einstein condensate (BEC).
The type of the nonlinearity is determined by the way the
condensate particles interact with each other.
For most dilute Bose systems it suffices
to consider only the Gross-Pitaevskii (GP) quartic non-linearity
which leads to the cubic \schrod equation (although, even for such systems
the beyond-GP approximations
are unavoidable in some cases \cite{Schick71}).
In general case, however, higher-order terms (which can account, for instance, for
multi-body interactions, self-energy effects, \textit{etc.})
can result in entirely
new physics as their infinite sum is an essentially non-perturbative
object with the features drastically different from what one
might expect from a perturbation theory \cite{Bel58}, an example to be given 
shortly after Eq. (\ref{e-eos}).

Thus, here we are going to view our $\Psi$
as a wave function of the 
effective BEC
described by the field operator $\hat\Psi$.
Then $\Psi$ can be considered as the expectation value
of the latter,
$
\langle \hat\Psi \rangle = \Psi
.
$
We assume that the full classical action 
can be decomposed into two parts (unless stated otherwise,
in this section we work in the high-energy units
$c=\hbar=1$):
\be\lb{e-act}
S = \widetilde S (\phi_i, \, \Psi) - \int {\cal V} (\Psi),
\ee
where the action $\widetilde S (\phi_i, \, \Psi) = \int \widetilde \lan$ 
and integration measure are
defined on some suitably chosen domain,
by $\phi_i$ we denote all other fields,
and
the potential energy density is defined as
\be\lb{e-pot1}
{\cal V} (\Psi) 
\equiv
\tfrac{1}{\Omega}
{ V} (\Psi)
=
\tfrac{1}{\beta \Omega} 
\left\{
\Omega |\Psi|^2
\left[
\ln{(\Omega |\Psi|^2)}
-1
\right]
+ 1
\right\}
,
\ee
up to an additive constant. 
Then at the ``classical'' level (replacing operators by their
expectation values) one of the  Euler-Lagrange equations can be always written
as
\be
\left[
\frac{\delta \widetilde S}
{\delta \Psi^*}
-
\int
\frac{d \, {\cal V} (\Psi) }
{d (|\Psi|^2)}
\Psi
\right]
\delta \Psi^*
=
0
,
\ee
which is equivalent to
\be
\frac{\delta \widetilde \lan}
{\delta \Psi^*}
-
\beta^{-1} \ln{(\Omega |\Psi|^2)}
\Psi
= 0
,
\ee
where by $\delta \widetilde \lan / \delta \Psi^*$ we loosely mean the 
functional derivative
of $\widetilde S$ 
with the integration dropped.
Thus, 
we readily recover the wave equation (\ref{e-xmain}) 
upon
a formal identification 
$ \hat{{\cal H}} \Psi \Leftrightarrow  \delta \widetilde \lan / \delta \Psi^* $.

Another way to see the fluidic features encoded in the logarithmic nonlinearity
is to look for solutions
of the  quantum 
wave equation in the Madelung form
\be\lb{e-anzapsi}
\Psi = \sqrt{\varrho}\, \text{e}^{i S}
, \ \
\varrho = \left| \Psi \right|^2
, \ \
\vec v = \frac{\hbar}{m} \vec \nabla S
=
\frac{\hbar}{i m} \vec \nabla \ln{(\Psi/|\Psi |)}
,
\ee
where $m$ is the \textit{inertial} mass of the condensate particle.
Then the wave equation
splits into two hydrodynamic ones - the equation of continuity
for the condensate particle density $\varrho (x)$
and the equation of  potential flow of superfluid for the velocity field 
$\vec v $
\cite{pn66,Dalfovo:1999zz}.
From the latter one immediately obtains that the zero-temperature (collisionless)
equation of state of the
logarithmic BEC in the first-order approximation is described by 
the Clapeyron-Mendeleev law,
\be\lb{e-eos}
p - p_0 = (m \beta)^{-1} \varrho + {\cal O} (\hbar^2) \propto T_\Psi \varrho ,
\ee
where $T_\Psi$ is in general a quantum (collisionless) kind 
of the  temperature conjugated to the 
information entropy, $S_\Psi \equiv - k_B \int |\Psi|^2 \ln{(\Omega |\Psi|^2)} d^3x$,
measuring the degree of spreading of a quantum object \cite{Zloshchastiev:2009zw},
see also Appendix.
For comparison, the corresponding
equation of state for the GP (quartic) condensate would be
$p \propto \varrho^2$, thus, the logarithmic Bose liquid is more ``ideal'' than the Gross-Pitaevskii one
yet non-trivial.
Therefore, the logarithmic condensate can 
be added to any microscopical
many-body system to serve as a calibrating background \cite{aakz}. 
This confirms the usefulness of the logarithmic nonlinearity
for describing the physical vacuum.
It is interesting also 
that since the Gross-Pitaevskii potential can be perturbatively derived from the 
logarithmic one by expanding near minima and cutting the infinite series
at the quartic term
we have found another example of how the essentially non-perturbative 
treatment, i.e., taking into account the infinite number of powers of $\varrho$, 
can drastically change the physical picture.

In this connection one can also mention that
the logarithmic  terms (usually of the form $\varrho^m \ln^n \!\varrho$) 
commence to appear in higher orders of
perturbation theory,
e.g.,
when taking into account certain combinations of  loop diagrams,
both in the relativistic scalar field theories \cite{Coleman:1973jx}
and 
condensed-matter Bose systems \cite{Schick71},
where $\varrho$ (modulo a dimensional scale factor) is the complex square of 
scalar field in the former case and the one of the condensate wave function in the latter.
This duality-type interplay between the \textit{relativistic} 
scalar field and \textit{non-relativistic} Bose liquids 
has a profound origin and will be discussed in more 
details later, in the section devoted to the BEC/spacetime correspondence.

To conclude this section, 
we have shown that one can mimic vacuum effects 
 by including the logarithmic nonlinearity
into the quantum wave equation or, alternatively,
by including into the full action the field
with the potential (\ref{e-pot1}).
If we view the nonlinearity
as a quantum gravity phenomenon 
then
we prefer to deliberately call the Bose-Einstein condensate virtual 
because it 
can not be physically separated 
from background and removed, in contrast to
its condensed-matter counterparts.
As a matter of fact, it {\it is} a background.

\scn{Mass generation}{sec-mas}

The exact form of the effective action $\widetilde S$ 
is unknown to us
but we can already guess the most obvious
of its features.
First, following the popular approach  
of taking into account vacuum effects by virtue of
introducing an auxiliary scalar field, 
see for example Ref. \cite{Graham:2002fw},
we can assume the psi-particle  to be described by scalar field.
At that, as long as here we are introducing this field as 
to account for the small
fluctuations of the BEC vacuum
and also we are going to describe
objects
with the quantum wave amplitude being much smaller
than the background value of the condensate wave function amplitude, 
the field-theoretical models can be constructed
in a covariant manner, for reasons which become clear below,
in the section devoted to the BEC/spacetime correspondence.
At the same time we have to keep this field non-linearized as to
account for the effects mentioned in the previous section.
In principle, since we are dealing with low-energy
effective models we are free to use any form of the covariant action
for the psi-field - 
as long as it is physically transparent, self-consistent, mathematically
manageable and
the corresponding field equation 
contains the logarithmic nonlinearity.
For instance, as to make the psi-particle field dynamical the
minimal
action must contain also the kinetic term
which must be quadratic otherwise 
no proper wave equation can appear.
Also, it is likely that $\widetilde S$ will contain
couplings of the psi-particle to other fields.
Thus, 
to get at least some idea about
how the conventional dynamical systems might be affected
by the logarithmic BEC vacuum,
in this section we are going to construct few toy models
complying with the above-mentioned requirements.
The issue of renormalizability of such models
is not a problem here because we do not require the Lorentz symmetry
to be exact at the length scales shorter than $\ell_0$, i.e.,
above the corresponding energy  and momentum  thresholds.
Then these critical values serve 
as the natural UV cutoff making the upper limits of momentum-space 
integrals finite and no UV divergences arises. 
The infrared divergences are not a problem either because in the low-energy
limit $E/E_0 \to 0$ the nontrivial structure of physical vacuum can be
neglected and one arrives at the relativistic models which are well-studied 
in this regard.

\sscn{Model with global symmetry breaking}{sec-tgl}

The simplest toy model is just the self-interacting one - involving
only the complex psi-field and no others.
While not having much of physical relevance on its own, it will serve
us as a good test-bed.
In $D$-dimensional spacetime
its Lagrangian can be written in the covariant form
\be
{\cal L} 
=
\ell_\Omega \,
\partial_\mu \psi \,
\partial^\mu \psi^*
-
{\cal V} (\psi)
,
\ee
where the potential is given by Eq. (\ref{e-pot1}); 
here and below the
factors like $\ell_\Omega$ are introduced for dimensionality reasons,
keeping in mind the original dimensionality of $\Psi$.

This model is invariant under a global change of phase of $\psi$
but
in the vacuum state the value of $\psi$ must be non-zero,
with a magnitude close to $1/\sqrt{\Omega}$ and arbitrary phase.
In other words, there is a degenerate family of vacuum states.
The latter circumstance together with the Goldstone theorem would suggest
the presence of the Nambu-Goldstone bosons in the theory.
To check this, we introduce the shifted real-valued fields $\varphi_1$
and $\varphi_2$:
\be\lb{e-shift}
\psi = 
\Omega^{-\frac{1}{2}}
+
\tfrac{1}{\sqrt{2 \ell_\Omega}}
(\varphi_1 + i \varphi_2)
,
\ee
and expand the potential near the minimum.
We obtain
\bw
\be
\lan  
=
\frac{1}{2}
\left[
(\partial \varphi_1)^2
+
(\partial \varphi_2)^2
\right]
-
\frac{1}{2} m_\psi^2 \varphi_1^2
-
\frac{\sqrt{2}}{\beta} 
\ell_\Omega^{(D-4)/2}
\varphi_1 (\varphi_1^2 + \varphi_2^2)
-
\frac{1}{4 \beta}
\ell_\Omega^{D-3}
(\varphi_1^2 + \varphi_2^2)^2
+
{\cal O} (\varphi^5)
,
\ee
\ew
where the quantity
\be
m_\psi 
=
2 
/\sqrt{ \ell_\Omega \beta}
\ee
can be viewed as the effective mass of 
the fluctuation of the logarithmic condensate (not 
to be confused with the mass $m$ of a bare condensate particle).
If 
the running behavior of $\beta$ 
turns out to be as derived in Ref. \cite{Zloshchastiev:2009zw},
\be\lb{e-betae}
\beta \sim (E_0 - E)^{-1}
,
\ee
then we expect
\be\lb{e-mpse}
m_\psi 
\sqrt{ \ell_\Omega}
\sim
\sqrt{E_0 - E}
,
\ee
i.e., its mass is not determined solely by the Planck scale:
for energy very small compared to $E_0$
it tends to the constant value, 
\be
m_\psi^{(0)} 
\equiv
m_\psi (E=0)
\sim
\sqrt{|E_0|/\ell_\Omega}
,
\ee
but at higher energies it alters thus reflecting the 
dynamical
nature of the physical vacuum.

Thus, in the broken symmetry regime this model describes
two kinds of particles, one massive and one massless.
The latter are the Nambu-Goldstone bosons
which describe the spatial variations
of the vacuum's phase.

\sscn{Model with gauge symmetry}{sec-tga}

Physically more useful toy model can be constructed
by coupling the condensate to the Abelian gauge field.
In $D$-dimensional spacetime its Lagrangian is
\be\lb{e-ablan}
\lan 
=
\ell_\Omega \,
D_\mu\psi^*D^\mu\psi 
-\tfrac{1}{4}F_{\mu\nu}F^{\mu\nu}
-{\cal V} (\psi)
,
\ee
with 
$
D_\mu=\partial_\mu+i e 
\ell_\Omega^{\frac{D-4}{2}}
A_\mu
$
and
$
F_{\mu\nu}=\partial_\mu A_\nu-\partial_\nu A_\mu
$, as per usual,
$e$ is the elementary electrical charge.

In general this Lagrangian is invariant
under the $U(1)$ local gauge transformation
and describes psi-particles and antiparticles
interacting with massless photons.
To see what happens in the regime of
spontaneously broken symmetry, we 
make again the shift (\ref{e-shift})
to eventually obtain
\be
\lan
=
\tfrac{1}{2}
(\partial \varphi_1)^2
-
\tfrac{1}{2} m_\psi^2 \varphi_1^2
-\tfrac{1}{4}F_{\mu\nu}F^{\mu\nu}
+
\tfrac{1}{2} m_\gamma^2 B_\mu B^\mu
+
\ldots
,
\ee
where
$B_\mu=
A_\mu+
\tfrac{1}{\sqrt{2} }
\ell_\Omega
e^{-1}
\partial_\mu \varphi_2
$
refers to the new gauge field
of the mass
\be\lb{e-mph}
m_\gamma 
=
\sqrt{2}
e
/\ell_\Omega
,
\ee
which does not run with energy.
We can see also that the masses of the photon and psi-particle and
the elementary charge 
are related by the formula
\be
\frac{e  m_\psi^2}{m_\gamma}
=
2^{3/2}/ \beta
\propto
E - E_0
,
\ee
which does not depends on $D$ or $\ell_\Omega$.
We remind that the Goldstone theorem is evaded
here because one of its prerequisites, the
Lorentz invariance, is violated in the
logarithmic theory as was shown
also in Ref. \cite{Zloshchastiev:2009zw}
in a different way.

Thus, we have established that the photon acquires
mass $m_\gamma$
and no massless Goldstone bosons appear.
The models support the Coleman-Weinberg idea
of the vacuum-induced spontaneous symmetry breaking \cite{Coleman:1973jx}
and
show that the possible
effect of the physical vacuum is that the photon
becomes massive.
Why its mass is so tiny small?
The clue is that the correlation length scale $\ell_\Omega$
can be very large -
in fact, as long as the parameter 
$\Omega = \ell_\Omega^{D-1}$ has the dimensionality
of the spatial volume and 
appears in the normalization condition
of the dimensionless wave function $\sqrt\Omega \Psi$,
it is tempting to
conjecture the cosmological-scale value for it, say, the volume of the
(observable part of the) Universe.
At least, 
that would  explain why the time-delay 
effects \cite{Collaborations:2009zq} are exactly as that small
as to
become visible precisely at the cosmological-scale distances.
Then, for  
the current value of $\ell_\Omega$ of about ten billion light years 
the above-mentioned characteristic masses can be estimated as
\be
m_\psi^{(0)} \sim 10^{-3} \div 10^{-2} \,  \text{eV},
\
m_\gamma \sim 10^{-35} \text{eV},
\ee
where for the former mass we imposed $E_0$ to be the Planck one
(which is valid if the external fields are weak enough as not to
 change the vacuum energy significantly).
These small yet non-vanishing masses indicate that their 
gravitational effect 
and contributions to the density of matter in the Universe
can be quite substantial, 
and can be phenomenologically estimated in the spirit
of the works \cite{Frieman:1991tu}.
Another thing that comes to mind
when looking at the formula (\ref{e-mph})
is that the appearance of $e$ therein
explains why it is the photon which mediates
the long-range interactions between the electrically charged 
elementary particles.
Recalling the analogy with superconductivity,
the photons in this model can be interpreted as the 
pairs of virtual particles and antiparticles, see also Ref. \cite{DesCloizeaux:1973qh}
and references therein.

\sscn{Other models}{sec-nab}

In our case, due to the interpretation of $\Psi$,
it suffices to represent
the complex-valued psi-field by two real scalars,
$\varphi_1$ and $\varphi_2$.
In general (for instance, when
the vacuum is required to be described by the multi-component Bose liquid), one may wish to consider
the multiplet of the scalar fields 
$\varphi^a$ which
belongs to a representation of the symmetry group $G$,
non-Abelian in general.
If the latter is spontaneously broken down to a subgroup
$H$ the fields acquire the non-zero expectation values $\varphi_0$.
Then the mass matrix for the gauge fields is
given by
$
(M^2_A)_{ab}=g^2 \varphi_0^\text{T} T_a T_b \varphi_0,
$
where $T_a$ are the group $G$'s generators,
$g$ is the gauge coupling constant.
The elements of $M_A^2$ which correspond
to the generators of $H$ vanish, therefore,
there appear $dim (H)$ massless gauge bosons
and $dim (G/H)$ massive ones.
The ``survived'' components of $\varphi$
acquire the mass
$
(M^2_\varphi)_{a b}= 
\left(
\frac{\partial^2 \ {\cal V} }{\partial\varphi^a\partial\varphi^b}
\right)_{\varphi=\varphi_0}
,
$
with ${\cal V}$ being the potential of the form (\ref{e-pot1}).

The fermions, such as neutrinos, can be also included into this picture as 
nothing prevents them from interacting
with the condensate.
Thus, they could also acquire mass, although the question
whether it would happen due to the condensate or due to the SM
Higgs boson remains open.

\scn{Topology and solitons}{sec-sol} 

The solitonic-type solutions of the logarithmic
wave equations 
have been known for a long time
\cite{BialynickiBirula:1979dp}.
However, at that time people were motivated by other things so they considered the potentials like (\ref{e-pot1}) ``upside down'', in which case no spontaneous symmetry breaking could arise.
It came as a surprise to us that nobody actually considered
other sector of the logarithmic
theory - the one where  
multiple topological sectors can in principle appear.
From the viewpoint of our theory, they were working 
with the ``Wick-dual'' theory -
in a sense that 
the two theories can be transformed into one another
either
by inverting the sign of $\beta$ or 
by the Wick-rotation
of an appropriate variable, as in the Euclidean 
field-theoretical approach \cite{Rajaraman:1982is}. 
The well-known example of theories related
by the Wick rotation is 
the quantum field theory at finite temperature $\beta^{-1}$ 
and the statistical mechanics on the $\IR^3 \times S^1$ manifold
with the $\beta$-periodic imaginary time.
In this connection, the relation between our $\beta$ and certain kind 
of non-classical
temperature
was outlined in Ref. \cite{Zloshchastiev:2009zw}, see also the Appendix.
Moreover, as long as $\beta^{-1}$ itself is shown there to be
proportional to $E - E_0$,
the natural energy of vacuum $E_0$ plays the role of the critical 
parameter at which a phase transition happens
(this can be seen from Eq. (\ref{e-mpse}) as well),
and the physical degrees of freedom
in each of the phases $E<E_0$ and $E > E_0$ can be very distinct.

As an example, let us consider one-dimensional
logarithmic \schrod equation.
In the dimensionless form it can be written as
\be
i \partial_t \psi + 
\left(
\partial^2_{x x} \pm \ln{|\psi|^2}
\right) \psi
=
0
,
\ee
where the plus (minus) sign 
corresponds to the theory 
with the potential (\ref{e-pot}) open downwards (upwards);
in practice this sign is associated with the sign of $\beta$.
For simplicity we impose the ansatz
$\psi = \exp{(-i \epsilon t)} \, \phi (x)$, 
with $\phi (x)$ being real-valued,
then the equation turns into the static one
(the moving solutions can be always generated 
by  performing the Galilean boost):
\be
\phi'' (x) 
-
d U_\pm 
(\phi)
/d \phi
=
0
,
\ee
where the potential is given by
\be
U_\pm 
(\phi) 
\equiv
\pm 
\tfrac{1}{2}
\phi^2 
\left(
1-
\ln{\phi^2} 
\right)
-
\tfrac{1}{2}
\epsilon 
\phi^2 
.
\ee

Let us consider first the ``plus'' case -
where the symmetry $\phi \to -\phi $ stays unbroken
because $\phi = 0$ is a stable local minimum of the potential
$U_+ 
(\phi)$.
The corresponding normalized solutions are called gaussons (on the BEC language they
would be called the \textit{bright solitons}):
\be
\phi_g (x)
=
\pi^{-1/4}
\text{e}^{-
(x - x_0)^2/2},
\ee
with the eigenvalue
$\epsilon = E_0 = 1 + \ln{\sqrt{\pi}}$.
Their stability is ensured by the integrability 
conditions because $E_0$ is the lowest 
bound for the energies of all possible normalizable solutions 
(generally referred as the BPS bound).

Now we turn to 
the ``minus'' case -
when 
the potential 
$U_- 
(\phi)$
 has two degenerate minima, at
$\phi = \pm \exp{(\epsilon/2)}$.
Therefore, one should expect
that all the non-singular and finite-energy
static solutions
can be cast into
four topological sectors,
according to the boundary conditions
\[
\text{e}^{-\epsilon/2}
[\phi(-\infty),\, \phi(\infty)] = 
\left\{
[-1,\, 1], \
[1,\, -1], \
[-1,\, -1], \
[1,\, 1]
\right\}
,
\]
and $\phi'(\pm\infty) = 0$.
The last two sectors contain the trivial solutions 
$\phi = - \exp{(\epsilon/2)}$ and $\phi = \exp{(\epsilon/2)}$, respectively, whereas
the former two contain the kink and anti-kink
solutions (\textit{dark solitons}, in BEC terminology),
with the non-vanishing topological charge.
The latter is 
defined simply as the difference of the topological indexes
\be
Q = \exp{(-\epsilon/2)} 
\left[
\phi(\infty) - \phi(-\infty)
\right].
\ee
To find the analytic form of the kink solution,
we solve the wave equation
with the above-mentioned boundary conditions.
We obtain the expression 
\be\lb{e-kink}
\int
\frac{d \phi}
{
\sqrt{\phi^2 
\left(
\ln{\phi^2} 
- \epsilon
-
1
\right)
+
\exp{\epsilon}
}
}
=
x-x_0
,
\ee
from which $\phi (x)$ can be found
after taking the indefinite integral.
Unfortunately, 
the 
latter
can not be expressed in known functions but
simple numerical analysis confirms that
Eq. (\ref{e-kink}) indeed represents the kink and anti-kink solutions.

Further generalizations are obvious, both in terms
of considering more dimensions and other symmetries.
If we relax the condition of real-valued $\phi (x)$
then the potential $U_- (\phi)$
takes the Mexican-hat shape
on the plane of the real and imaginary components of $\phi$.
The topological classification is usually based 
on the homotopy groups $\pi_n (S_m)$ \cite{Steen}.
For instance, the homotopy
group for the Abelian model (\ref{e-ablan})
at $D=3+1$ is $\pi_2 (S_1) = 0$, i.e., no nontrivial
homotopy sectors of solutions can exist
whereas 
at $D=2+1$ its homotopy group is $\pi_1 (S_1)$ which is a
winding number group.
The latter implies that 
in principle
in effectively 
$(2+1)$-dimensional Abelian gauge models with 
the condensate 
the magnetic flow becomes quantized
and the vortex solutions can appear 
\cite{Abrikosov:1956sx}.

\scn{
BEC Vacuum vs. curved spacetime}{sec-bvc}

Now, as long as the (quantum) gravity is concerned,
how  can one reconcile the BEC description 
of the physical vacuum 
with the concept
of curved spacetime which 
is traditionally being used for describing the gravitational interaction?

\sscn{Emergent spacetime}{sec-ret}
Let us first
recall that in majority of physically meaningful cases
one can establish a formal correspondence
between the inviscid Bose liquids and  manifolds of non-vanishing
Riemann curvature.
For instance, the following fluid/gravity
correspondence is well-known
\cite{Unruh:1980cg}:
the propagation of small perturbations inside an
inviscid
irrotational barotropic fluid,
characterized by the background values of the density $\varrho$, 
pressure $p$ and velocity $\vec v$,
is analogous to propagation of
test particles along the geodesics
of the 
pseudo-Riemannian manifold with
the 
metric 
\be\lb{e-metr1}
g_{\mu\nu} 
\propto
\frac{\varrho}{c_s}
\left[
\baa{ccc}
-(c_s^2-\vec v^2)& \vdots & -\vec v \\
\cdots\cdots & \cdot & \cdots \\
- \vec v & \vdots & \textbf{I}
\eaa
\right]
,
\ee
where 
$c_s = \sqrt{\partial p /\partial \varrho} $ 
is the speed of ``sound'' - the propagation speed of wave-like fluid fluctuations.
This metric tensor is defined up to a constant factor
which value is determined by measurement units and boundary conditions.
Notice  that while inside the background fluid
the notions of space and time are clearly separated
(such that one can assume the fluid being non-relativistic),
the small perturbations themselves couple to the 
metric which treats space and time in a
unified way. 
If we treat such fluid as a non-removable 
background then
this metric describes the induced spacetime geometry.
The latter
should not be confused with the relativistic gravitational effect of 
the ideal fluid as a source
introduced via stress-energy tensor in the 
Einstein field equations (EFE).
Instead, 
as long as the physical vacuum is concerned,
for a given metric (\ref{e-metr1}) 
one can always define the induced matter stress-energy tensor
\be\lb{e-sete}
T_{\mu\nu}^{(\text{ind})} 
\equiv
\kappa^{-1}
\left[
R_{\mu\nu} (g) - \frac{1}{2} g_{\mu\nu} R (g)
\right]
,
\ee
thus, EFE are interpreted here not as the differential 
equations for the unknown metric but
rather as an expression for the  stress-energy tensor of
the effective matter to which the small fluctuations and test particles couple.
If an observer operates only with such fluctuations  then 
this is the only matter s/he is going to ``see'' directly.
Macroscopic (composite, finite-size) bodies also couple to the induced metric if they
consist of the elementary particles which do not
violate the small-fluctuation condition - such that
the overall density is much less than the critical one.

Using Eq. (\ref{e-anzapsi}) one can show that for
the generic bulk Bose condensate 
described by the \textit{non-relativistic} quantum wave equation
\be\lb{e-becgeneq}
\left[
- i \hbar \, \partial_t
- \frac{\hbar^2}{2 m} \vec \nabla^2
+
V_\text{ext} (\vec x,\, t)
+
F(|\Psi|^2)
\right]
\Psi
= 0,
\ee
the zero-temperature equation of state and hence
velocity $c_s$ can be determined from the
differential equation
\be
m \vec\nabla p - 
|\Psi|^{2}  \vec\nabla F =
{\cal O} (\hbar^2)
,
\ee
the square
of the BEC wave function yields the condensate density, as usual.
By solving this equation we obtain
\ba
&&
p - p_0
=
m^{-1} \int\limits_0^{|\Psi |^2} 
\varrho
F'(\varrho) d \varrho
+
{\cal O} (\hbar^2)
,
\lb{e-eosdiff}\\&&
c_s^2 
\equiv \partial p /\partial (|\Psi|^2)
=
m^{-1} |\Psi |^2 F'(|\Psi |^2),
\ea
and
the induced metric tensor takes the form:
\bw
\be\lb{e-metr2}
g_{\mu\nu} 
\propto
\frac{|\Psi|}{\sqrt{F'(|\Psi |^2)}}
\left[
\baa{ccc}
-
\frac{1}{m}
|\Psi |^2 F'(|\Psi |^2)
-
\frac{\hbar^2}{m^2} \left[\vec\nabla \ln{(\Psi/|\Psi |)}\right]^2
& 
\vdots & 
\frac{i \hbar}{m}  \vec\nabla \ln{(\Psi/|\Psi |)} \\
\cdots\cdots & \cdot & \cdots \\
\frac{i \hbar}{m}  \vec\nabla \ln{(\Psi/|\Psi |)}  & 
\vdots & \textbf{I}
\eaa
\right]
.
\ee
\ew
The value $c_s$
thus becomes the maximum attainable propagation velocity of any object
whose quantum wave amplitude is much smaller
than the  magnitude  of the background condensate wave function.
For instance,
when assuming the logarithmic condensate,
$F (\varrho) \equiv \beta^{-1} \ln{(\Omega \varrho)}$, 
then in absence of any additional matter
Eq. (\ref{e-eosdiff}) yields Eq. (\ref{e-eos}) from which
we obtain  
\be\lb{e-csbeta}
c_s =  1/\sqrt{m \beta}
\equiv
c_\beta 
,
\ee
provided $m \beta > 0$.
One can immediately see that the maximal propagation
velocity of small excitations in the long-wavelength approximation
does not depend on density
which makes the logarithmic BEC distinguished among other
kinds of condensates.
From last formula one can  derive also the second Einstein's postulate:  
if we recall Eq. (\ref{e-betae}) and assume an absence of extra fields
so we can choose the proper BEC energy $|E_0| = m c^2$,
with $c$ playing the role of the units conversion factor,
in the leading approximation.
Then we indeed arrive at the fundamental velocity constant:
$
c_\beta \leqslant \sqrt {|E_0| / m} \leqslant c
$.

Thus, in the BEC-vacuum approach the relativity is an emergent rather than
a fundamental phenomenon\footnote{The question whether the 
general relativity is an effective theory
has been raised long time ago \cite{Sakharov:1967pk}. 
Also,
the early attempts to describe the physical vacuum as 
superfluid were dated as far back as 70's \cite{Sinha:1976dw} (however,
neither there nor in later works \cite{Novello:2002qg} 
any specific wave equations for the physical vacuum's wavefunction 
were proposed,
to our best knowledge, and the debates about a
specific expression for the vacuum energy density still continue).},
EFE and dependent concepts do not have any
fundamental meaning on their own
but rather represent an
approximate long-wavelength  description
valid only within certain energy and length scale
(after all, the Lorentzian geometry is what it is - 
a way of measuring distances, and the gravitational ``field'' in
general relativity is known
for not possessing a proper stress-energy tensor). 
In fact, 
some predicted quantum gravitational phenomena, such as the Hawking radiation,
can be derived without the use of EFE \cite{Hawking}
whereas others, such as gravitons and
gravitational waves (at least, in current 
formulation), strongly rely upon EFE, and therefore, a careful treatment
is needed there.
The BEC-vacuum description of the black holes is also slightly different
from general relativistic:
while the analogue spacetimes may possess event horizons
it is only  long-wavelength excitations which follow geodesics
and thus it is only them which might experience the irreversible
properties of horizons.
If a measuring apparatus operates with the objects which somehow
do not satisfy the small-amplitude and long-wavelength conditions then no ``canonical'' event horizons
can be detected.
The space-time singularities, i.e., the points where
the Riemann tensor computed from the induced metric
diverges,
can not be attributed
to reality as  the small-amplitude and long-wavelength
conditions, main prerequisites of the induced relativity,
are strongly violated there.
It should be remembered also that due to the
original
relativistic time coordinate
being synchronized
with the Newtonian time
the BEC-induced geometries
automatically fulfill
the requirement of \textit{stable causality} which prevents the 
appearance of the causal pathologies the general relativity suffers from.

Numerous examples of the fluid-gravity isomorphisms
and further discussions
can be found in the books \cite{Novello:2002qg}.
In particular, the BEC-gravity analogue models 
have been
already studied in Refs. \cite{Garay:1999sk}, although they
dealt with condensed-matter systems without referring
to the physical vacuum and mass generation mechanism,
an extensive bibliography
can be found in Refs. \cite{Novello:2002qg,Barcelo:2005fc}.
Moreover, the nonlinear wave equations in those models
are not of the logarithmic type,
therefore, 
they do not
possess the above-mentioned Planck relation, energy additivity 
and constancy of $c_s$ properties
jointly
which makes them
less suitable for describing the fundamental background.

On a practical side, the BEC-gravity analogy\footnote{In our case the term 
BEC/spacetime correspondence or duality would be more appropriate 
provided 
we assume the broader meaning
of the condensate as the coherent ground state of superfluid described by a single
wavefunction. In general, however, the notion of superfluid is more broad and complex
than that of BEC.
} means that the (physical) observer operating at the length scale larger than the size of 
elementary
fluid elements of quantum Bose liquid (which is of order $\ell_0$)
is not able to distinguish the propagation of small 
fluctuations in the fluid from the geodesic motion
of test particles on an appropriately chosen manifold.
To resolve the underlying microscopic structure of 
the liquid s/he has to input therein energy sufficient to reach the critical value $|E_0|$
which corresponds to the length resolution $\ell_0$.
Then, as mentioned in previous section, the system ``jumps''
into other phase, with different physical degrees of freedom,
the process which resembles the transition between the phonon and free-particle phases
in Bose gases.
But otherwise  these two descriptions,
Bose-liquid and geometrical one, are dual - equivalent and equally ``effective''
(and may be not the only possible),
and the choice between them is purely a matter
of taste and/or practicality.
For example,
while the simple superfluids (irrotational, barotropic, one-component) can be associated 
with simple pseudo-Riemannian manifolds (real, four-dimensional, torsion-free, metric-compatible, \textit{etc.}) -
such that one can employ the whole machinery of the Riemann
geometry,
the geometrical description of
the liquids with any of the above-mentioned restrictions
relaxed 
can easily go beyond
the Riemann geometry
and
become complicated and/or physically 
non-transparent \cite{Visser:1997ux}.
Besides, the applicability of either description depends on
a concrete physical problem it is applied to.
It seems that
the propagation of test point-like objects
inside the physical vacuum is more conveniently
described within the framework of the relativistic approach 
(although, some corrections apply \cite{Zloshchastiev:2009zw})
but the strong quantum processes such as
the phase transitions related to
the formation or depletion of the Bose liquids (``spacetimes'') themselves
can be
mathematically 
formulated
only
at the underlying BEC level (in fact,
within the framework of the Lorentzian geometry such phenomena
can not be even well-defined).

Thus, the BEC/spacetime correspondence allows to look at some old problems
at a different angle of view.

\sscn{Mach's principle and locality}{sec-mpr}
In its most popular formulation the Mach's principle states that
the local inertial properties such as mass
are determined by the
total mass distribution in the Universe.
While Einstein himself had this  in mind
when constructing  general relativity
the latter does not comply with the Mach's principle
favoring instead the strong equivalence one.
The attempt of fixing that 
without breaking general covariance
has been made in the
theories of 
scalar-tensor gravity \cite{Sciama:1953zz}.
In those approaches the Mach's principle is partially taken into account  
by making the gravitational constant a dynamical variable, at the cost of
postulating the additional  field - the scalar one.
The origin of this hypothetical scalar remains unclear so far, 
moreover, being Lorentz-covariant the scalar-tensor gravitational models
do not address the following two locality issues.

If the physical vacuum is trivial then
an observer in the otherwise empty space   
would not be able to determine whether s/he has any inertia - 
due to the absence of any reference frame.
The latter can be 
immediately created once a probe object appears
somewhere
else.
Therefore, the observer 
is supposed to instantaneously find out own inertial
properties with respect to that frame,
no matter how far the probe is located or how ``massive'' it is. 
Another locality issue which arises in a theory with
the trivial
vacuum is the following:
if we talk about interacting systems
in general
then
what do we mean by energy of interaction, 
how can we differentiate
``interacting'' and ``non-interacting'' systems, 
how does a system
``know'' about the form of the potential it is supposed to obey
when interacting with 
other system(s)?

To address all these questions in our approach,
let us recall that
the  BEC vacuum  is an essentially quantum object yet
its correlation length $\ell_\Omega$ can have the cosmological-scale 
value, as mentioned above,
and
the properties of its fluctuations
are obviously determined by the whole matter distribution 
in the Universe.
As a matter of fact, the condensate gives rise to masses of
particles in a way similar to the gap generation mechanism
in superconductors, as we have shown earlier.
Therefore, the nontrivial vacuum can naturally
serve as the physical realization of the Mach's principle:
it  introduces the universal frame of reference 
and gives meaning to 
the ``action-at-a-distance'' processes in general and to the inertia in particular.
In this framework
the Lorentz-covariant models involving the global scalar field
is just a way to account for the BEC vacuum's
effects in
the (approximate) relativistic 
manner 
only for length scales larger than $\ell_0$ 
and for transfer 
energies 
below $|E_0|$ -
similarly to what we have done in Sec. \ref{sec-mas}.
At that, one does not need to introduce any kind
of ``gravity'' in the Euclidean space because the only mass
parameter there, $m$, is the \textit{inertial} mass 
of the condensate particle.

The issue of how to unambiguously define the concept of
interaction under the conditions of the strong long-range correlations 
is resolved in the logarithmic BEC vacuum
due to the above-mentioned energy
additivity property which is preserved in the logarithmic quantum mechanics:
an interacting energy of any two systems described by wave functions $\Psi_1$ 
and $\Psi_2$ (when taken separately from each other) still
can be  defined
as the difference 
$E (\Psi) - E (\Psi_1) - E (\Psi_2)$
where $\Psi$ is the wave function of the whole composite system.
This definition naturally incorporates the quantum-mechanical nature
of interactions: it preserves the notion of non-interacting
systems whereas the interaction energy defined in such way
is a measure
of how much does the overall state vector $|\Psi \rangle$ differ from the 
plain product $| \Psi_1 \rangle \otimes |\Psi_2 \rangle $.

\sscn{Cosmology}{sec-cosm}
According to  current cosmological paradigm, the early Universe's 
large-scale structure
had a phase of the exponential expansion (inflation) followed
by the reheating and, subsequently,  
radiation- and matter-dominated phases \cite{Guth:1980zm}.   
It is believed that without introducing the inflationary phase 
it would be difficult to
explain the horizon, flatness and monopole problems.
For the role of the agent driving the inflation
one usually appoints the global scalar field called the inflaton
and considers some kind of the scalar-tensor gravity rather
than the original Einstein's theory.

Despite the overall success and popularity of the scalar-driven
cosmological models, few questions remain unanswered.
The main one is what is the physical nature of the inflaton,
in particular,
why did it appear in the early Universe 
before any other fields and particles we know so far,
why its current vacuum expectation value is the way it is,
why the current expectation value of its potential energy,
known also as the (effective) cosmological constant,
is so extremely small yet nonzero in present epoch.
On top of that, if one associates this effective cosmological constant 
with the vacuum energy then one immediately arrives at the
above-mentioned cosmological constant problem \cite{Zeldovich:1967gd}.
How would all these problems look from the viewpoint of the cosmology
incorporating the BEC-vacuum idea?

First thing to notice, 
the notion of the cosmological constant makes sense in a relativistic 
theory only, therefore,
within the framework of the BEC  approach
this constant can refer at most to the energy of  small
fluctuations of the vacuum
above a background value but not to the energy of vacuum itself \cite{Volovik:2006bh}.
Thus, in the BEC-vacuum cosmology this constant does not 
have any fundamental physical meaning and the related problems simply
do not occur in first place. 
 
Second, 
if typical 
energies of density fluctuations and masses of elementary particles
are less than $E_0$
then
the vacuum stays in the BEC phase 
and
the Lorentz-symmetric cosmological models based
on the spacetime metric tensor
and scalar fields are obviously a good approximation, therefore,
the physical conclusions based on the standard 
Friedmann-Lema\^itre-Robertson-Walker (FLRW) models 
remain unaltered.
Moreover, in the BEC phase many of the conclusions based
on scalar-driven models
remain unaltered as well, as long as one 
adopts a suitable form of the scalar-tensor field-theoretical action. 
However, in the close vicinity of the threshold
the relativistic description begins to fail:
of course,  
as one approaches
more and more early stages of the Universe's evolution, 
one can still employ the relativistic fields but the price
will be that this description will become more and more ``effective'' and 
less and less natural. 
In practice this means that one will need to adjust 
the form of the covariant field-theoretical action at 
each range of energy scale by hand.
 
Finally, let us discuss
the problems which led to the inflation proposal and
give them explanations based on the BEC-vacuum idea: 

\begin{itemize}
\item The monopole problem is eliminated in the BEC-vacuum  cosmology  
for the above-mentioned reasons: the stable GUT monopoles
predicted so far  are the solutions of relativistic
field equations possessing a large mass.
The latter circumstance violates
the requirements for the BEC/spacetime correspondence's
validity and thus 
the relativistic monopole production 
in the early Universe
is hardly justified  even on theoretical grounds.
\item The flatness or  cosmological fine-tuning problem
was motivated by the analysis of the
Friedmann equations which are again intrinsically relativistic,
therefore, they can not be extrapolated to arbitrary short length scales
and the genuine evolution of the density of matter and energy 
in the Universe did not have to obey them all the time.
The reason why the density is so close to the critical one  
is that just an instant before the vacuum BEC was formed and 
its
fluctuations became small enough there was
no concept of curved Lorentzian spacetime available yet.
Therefore, 
at that moment
the total density had a 
critical value (corresponding to the flat space)
and its large-scale average value could not change much since then - 
provided the BEC does not rarefy much. 
The latter can be achieved by self-sustainability due to nonlinear effects \cite{aakz},
some sort of trapping potential, 
and/or boundary conditions for the wave equation the background BEC obeys.
At that, one should not confuse, for instance, the spacetime (Hubble)
expansion as viewed
by the internal observer operating in the small-perturbation regime
with the dynamics of the BEC background itself:
below we demonstrate certain physical setup in which
the BEC background
flows with constant velocity 
(if viewed as an embedding in the fictitious Euclidean space) 
while the
observer sees herself inside the FLRW-type universe.
\item The 
problem of reconciling the early-Universe
cosmology with the second law of 
thermodynamics which
is closely related to the horizon problem (homogeneity and isotropy)
and
leads either to the inflation proposal
or to the Weyl curvature hypothesis 
\cite{Penrose:1980ge}
can be reformulated in the BEC-vacuum cosmology as follows.
During some epoch of the very early Universe when any conventional matter was absent the
large-scale evolution was determined mainly 
by the vacuum, logarithmic condensate.
The Weyl curvature hypothesis requires then that 
the induced metric (\ref{e-metr2}) must be conformally flat
during that epoch.
Below we show that it is indeed the case.  
The horizon problem can be thus explained by 
the macroscopic size of the essentially quantum vacuum -
as long as the latter 
is viewed as the BEC embedded into the Euclidean space with absolute time
such that its particles tend to occupy the lowest state
and any quantum exchanges happen instantly. 
In the case of the BEC-vacuum cosmology the
correlation length $\ell_\Omega$
can be interpreted as the size of 
the observable part of the Universe.
Indeed,
as long as an observer usually operates with the probe objects, 
such as photons and other elementary particles
with energies less than $E_0$,
s/he
is bound to the relativistic regime and
thus unable to probe not only the distances smaller than $\ell_0$
but also larger than $\ell_\Omega$.
This also means that
the regions relativistically disconnected from us  
can nevertheless  affect our Universe - e.g., by virtue of the large-amplitude 
density fluctuations  for which $\delta (|\Psi|^2) \not\!\ll |\Psi|^2$.
This fits the long-discussed idea of our Universe being 
a patch inside the much ``larger''
region, called the Multiverse,
which may explain the
dipole anisotropy of the cosmic microwave background
and
coherent large-scale flow of galaxy clusters
\cite{Turner:1991dn}.
Besides, since the Multiverse can contain many patches  with
different BEC vacua (or currents, if viewed
as the Euclidean embeddings)
which separately nucleated 
during the Bose condensation epoch,
the chaotic inflationary scenarios \cite{Linde:1986fd}
are compatible with the BEC-vacuum cosmology as well.
\item The long-standing problem
of how to formulate the early-Universe cosmology on quantum-mechanical grounds,
commonly referred as the \textit{quantum cosmology},
is treated in the following way:
as long as the Lorentzian geometry is the induced effective phenomenon
valid only for certain scales of length and energy,
the metric tensor does not need to be quantized {\it per se}
otherwise it leads to the double-counting similar to the one
which appears when one attempts to (re)quantize phonons \cite{Volovik:2006bh}.
What happens actually to be  quantum is the underlying background BEC vacuum, its 
ground-state
wave function 
induces nontrivial geometry by virtue of the map (\ref{e-metr2}).
The effective metric thus emerges as one of the low-energy collective
modes of the vacuum.
Further, the metric defines the stress-energy tensor (\ref{e-sete}) which in turn determines
the large-scale evolution of the Universe as well as
the distribution of matter therein.
The other SM-type interactions, chiral fermions and gauge fields,
emerge as well - as the different quasi-particle excitations of the quantum vacuum liquid
(not to be confused with the bare particles of the latter),
similarly to the mechanisms proposed in a theory of condensed matter \cite{Sinha:1976dw,Novello:2002qg},
although it might require adding the Fermi component
to the Bose liquid describing the physical vacuum.
\end{itemize}

To give an analytical illustration of these statements,
we consider the following physical setup
which is the simplest one can imagine of yet
can be realized in the ``early'' Universe at some stage:
the just-formed BEC vacuum described by the logarithmic condensate
is the predominating form of matter, any other kinds 
have not appeared yet.
Then the induced metric (\ref{e-metr2})
is completely determined by a solution of Eq. (\ref{e-becgeneq}) with 
$F (x) \equiv \beta^{-1} \ln{(\Omega x)}$,
namely
\be\lb{e-beceqlog}
\left[
- i \hbar \, \partial_t
- \frac{\hbar^2}{2 m} \vec \nabla^2
+ V_\text{ext} (\vec x,\, t)
+
\beta^{-1} \ln{(\Omega |\Psi|^2)}
\right]
\Psi
= 0
,
\ee
under certain boundary conditions.
While both those conditions and the trapping potential 
are still
unknown to us, one can
already deduce a very important general feature:
due to the separability property
of the logarithmic \schrod equation
its simplest ground-state solutions 
have the phase which is
linear with respect to the radius-vector \cite{BialynickiBirula:1976zp},
\be\lb{e-phaselin}
i \ln{(\Psi_0 (\vec x, t)/|\Psi_0 (\vec x, t)|)} 
\propto 
\vec v^{(0)} \cdot \vec x + f (t)
,
\ee
which indicates, upon recalling Eq. (\ref{e-anzapsi}), 
that the background condensate flows with a constant 
velocity $\vec v^{(0)}$ if viewed as an embedding into the Euclidean space.
Together with Eq. (\ref{e-csbeta}) it means that 
the
geometry induced by such {\it panta rhei} solutions is conformally flat,
\be\lb{e-metlog}
d s^{2}_{(\beta)} \propto
\Omega
|\Psi_0 (\vec x, t)|^2
\left[
- 
c_\beta^2
d t^2
+
(d \vec x - \vec v^{(0)} d t)^2
\right]
.
\ee
At the level of metric, the value of $\vec v^{(0)}$ becomes irrelevant and
can be set to zero by an appropriate
coordinate transformation;
at the level of the Euclidean observer
this corresponds to selecting the Galilean frame
of reference comoving with the background.
Obviously, for manifolds with
such metrics the Weyl tensor vanishes 
so they are of type $\textbf{O}$ in the
Petrov classification \cite{Petrov:2000bs}.
This is the class where all the FLRW spacetimes, including those
expanding with an acceleration, belong to 
(in general relativity the manifolds corresponding to isolated gravitating objects belong to  type
$\textbf{D}$, spacetimes of other types involve gravitational waves of different kinds).
Therefore, 
for our physical setup we will necessarily obtain
one or another
family of the
FLRW spacetimes - just written in the conformally-flat coordinates, like in the
kinematic cosmology \cite{CFS}.

Further, to derive
the  induced stress-energy tensor
corresponding to our setup
we use the definition (\ref{e-sete})
where assume that the metric is given by last equation,
$g_{\mu\nu} d x^\mu d x^\nu = d s^{2}_{(\beta)}$.
With the help of the conformal rescaling technique
we immediately obtain
\bw
\be\lb{e-setelog}
\kappa T_{\mu\nu}^
{(\beta)}
=
\tilde D
\left[
\nabla_\mu \nabla_\nu \Phi
-
\nabla_\mu \Phi \nabla_\nu \Phi
-
g_{\mu\nu}
\left(
\nabla_\lambda \nabla^\lambda \Phi
+
\tfrac{1}{2}
(\tilde D-1) 
\nabla_\lambda \Phi \nabla^\lambda \Phi
\right)
\right]
,
\ee
\ew
where 
$\tilde D \equiv D -2 = 2$,
$\nabla$ is the covariant derivative
with respect to the metric $g$,
and
we have designated
\be\lb{e-phipsi}
\Phi \equiv \ln{
\left( 
\Omega |\Psi_0 (\vec x, t)|^2
\right)
} 
,
\ee
up to an additive constant.
This stress-energy tensor strongly resembles 
the one of
some theory with scalar field, 
and indeed, one can check that
it
can  be  formally derived, by varying the metric,
from the following scalar-tensor
gravity action functional
\be
S^{(\beta)} [g,\, \not\!\Phi]
\propto
\int d^D x
\sqrt{- g}
\,
\text{e}^{\tilde D \Phi}
\left[
R + \tilde D (\tilde D + 1) 
(\nabla \Phi)^2
\right]
,
\ee
where the notation ``$ \not\!\Phi $'' reminds 
that
the ``dilaton field'' $\Phi$ has been already fixed by the solution 
of the BEC quantum wave equation, in this case it would be Eq. (\ref{e-beceqlog}).
Being entirely formal and analogous (because in reality 
both the metric and ``dilaton'' are 
determined
by the BEC vacuum which is in the state described by $\Psi_0 (\vec x, t)$),
this action nevertheless confirms what was written before about 
fundamental scalar field:  
it explains why the relativistic models involving
scalars, such as the scalar-tensor gravity
or (bosonic sector of) supergravity,
yield the expressions for metric tensors
which
seem to
provide the good qualitative description of the large-scale
evolution of the early Universe 
and agreement with current observational data 
yet no scalar partner has been detected.
Moreover, this duality between the non-relativistic quantum BEC equation and
relativistic classical
scalar-tensor gravity
also shows the already discussed
limitations of the relativistic description
alone:
once the BEC vacuum goes into the different quantum state
represented by other solution of Eq. (\ref{e-beceqlog})
one gets a different expression for the induced metric
and, therefore, for the induced stress-energy tensor and covariant
action.
In fact, for more complicated physical setups even the condition (\ref{e-phaselin}) 
leading to conformal flatness  
can be relaxed to the asymptotic one.
Therefore, depending on a physical background (determined by external potential
and boundary conditions) and the quantum state the vacuum stays in,
the small fluctuations and test particles
obey several covariant actions.
The unified picture can be seen only at the level
of the quantum wave equation for the background BEC.

To conclude, in this section we have shown that 
relativistic gravity can be viewed as the  phenomenon
which emerges due to the long-wavelength fluctuations of the quantum yet macroscopical 
object, the non-trivial BEC vacuum.
In fact, it can be useful to think in terms of the duality
rooted in some kind of uncertainty principle:
one can view the physical vacuum
either
as the Lorentzian spacetime 
(which, as we know, can have the non-vanishing
Riemann curvature but no well-defined microscopical structure)
or
as the flat Euclidean space, along with the Newtonian time parameter,
filled with
some kind of background quantum liquid (such that the microscopical structure 
is well-defined but no curved-spacetime description is possible).

\scn{Conclusions}{sec-cnc}
It is shown that on the language of field theory
the logarithmic nonlinear quantum wave equation
can be  interpreted
in terms of
the background Bose-Einstein condensate by analogy with the Bogoliubov-Ginzburg-Landau 
theory \cite{Bogoliubov47}.
Recall that the latter is known as the effective
mean-field
theory of superconductivity
which not 
only helped to figure out most of phenomenological
implications long before the underlying
microscopical model was formally written down
\cite{Bardeen:1957kj}
but also served as a guiding light on
a crooked path of the theoretical constructing
of the BCS theory. 
In our case the microscopical theory of the background BEC 
can be regarded as the quantum gravity itself
so there is a hope that the non-axiomatic approach based
on logarithmic wave equation will
do its job here as well.
As for the underlying microscopical theory then
the presence of two length scales, 
$\ell_0$ and $\ell_\Omega$,
points out
at the possibility that the noncommutative-space extension of quantum mechanics
(NCQM) is a strong candidate -
and, indeed,
the objects which resemble the Cooper pairs 
(and can be viewed as the dipole-order approximation
of a fluid element) do arise 
there naturally \cite{Bigatti:1999iz,ncqmz}.
Another approach would be to leave the spatial commutators
intact but instead treat the (bare) condensate particles and Euclidean space
as the underlying entities, and construct the microscopical theory
in the spirit of the conventional non-relativistic theories of superfluidity and superconductivity,
and then use the maps like (\ref{e-metr2}) and (\ref{e-phipsi})
to translate the results into the language of a physical (relativistic) observer.
In any case,
once the vacuum liquid is formed it can be regarded as the most fundamental
object (due to its ground state being described by a single wave function only)
whereas
the particles and interactions observed by a physical observer
are represented by its different
modes - collective ones and excitations.

It is worth mentioning also
that since
the quantum gravity is concerned
there exists 
the conceptual difference between the 
interpretation of our
Bose-Einstein condensate 
and its condensed-matter counterparts:
unlike the latter it
represents the fundamental (non-removable) background.
This essentially implies that not only the
objects which are being observed 
are being immersed into the condensate
but also are the observers themselves with
their measuring apparatus.
Thus, such condensate affects not only the ``objective'' motion of particles
but also the process of measurement itself which results
in the nonlinear
corrections to the quantum wave equation, see some discussions in the Appendix and references
therein.
That is why the theory with the logarithmic
nonlinearity \cite{Zloshchastiev:2009zw} can be also viewed as 
(the nonlinear extension of) quantum mechanics \cite{BialynickiBirula:1976zp,cza98}.
The latter is believed by many to be
the consistent way of handling 
the difficult places of the conventional quantum
mechanics - such as the measurement 
problem 
(wave-function collapse {\it vs} many-worlds interpretation)
\cite{Singh:2007dj}.

Further, we
demonstrated  that this kind of nonlinearity 
can cause in principle
the spontaneous symmetry breaking 
and mass generation phenomena. 
The mass generation mechanism based on vacuum fluctuations
is universal in a
sense that it may supplement the electroweak one
(by generating the masses of the photon and Higgs boson, for instance) but also
it is capable of enhancing or even replacing the latter, under
certain physical circumstances.
The role of BEC seems to be natural here 
because the mass generation by such a highly non-classical object  
naturally
serves as a physical realization of
the Mach's principle.
We proposed few toy models 
to estimate the values of the generated masses of the otherwise
massless particles such as the photon.
We wrote those models in a covariant form 
and also the above-mentioned 
effect of the vacuum upon the measurement procedure
is neglected as well.
These assumptions seem to be a
good approximation when one works in the energy range below the
vacuum energy threshold $E_0$
and, therefore, deals with small perturbations
of the vacuum and elementary particles being also small fluctuations.

The straightforward computation shows that
the photon mass, gained due to its interaction
with the quantum-gravitational vacuum represented
by the logarithmic condensate,
can be expressed
as a ratio of the elementary electrical charge
and the length related to one of
the parameters of nonlinearity.
We gave some phenomenological arguments for why this (coherent) length's scale
can be related to the size of the (causally connected part of) Universe
as well as why the electric charge appeared in the formula.
It once again confirms the choice of the wave equation's nonlinearity
to be of the logarithmic type.

The relation of the BEC description of the physical vacuum to the
curved-spacetime one is established 
via the well-known fluid-gravity correspondence.
The latter presumes the introduction of two types of
observers - physical or relativistic, operating in the long-wavelength
excitations regime, and mathematical or absolute one, acting in the fictitious Euclidean space.
The latter is essentially unobservable yet allows to formulate certain phenomena
in a more consistent way.
The dictionary between the languages ``spoken'' by these two observers
is still mostly unknown but it is already started:
we formulate the map which relates
the long-wavelength solutions of the non-relativistic nonlinear quantum wave equation to
the metric manifolds arising in the 
relativistic classical
scalar-tensor gravity.

To reconcile this description also with the current 
cosmological paradigm 
we advocated the idea that
the curved-spacetime description of the Universe's large-scale evolution is valid only
in the long-wavelength approximation, and it is not the only possible or most convenient:
one can also describe it (hydro-)dynamically
as the Bose liquid  
which flows in certain way
when viewed as an embedding into the Euclidean space.
Such description allows to take a different look at some long-standing problems of both the
standard and inflationary cosmologies, and also hints at the possible ways
of formulation the theory of quantum gravity.
In particular, the generation of FLRW models and scalar field
is shown explicitly for the physical setup involving
the logarithmic BEC in the long-wavelength approximation.

Finally, the 
generic topological properties and
corresponding solitonic solutions of the theories
with ``logarithmic'' condensates 
related by the Wick rotation
(or, alternatively, by inversion of the sign of the parameter $\beta$)
were compared and discussed.
The role of the natural energy of vacuum 
as a critical parameter for certain phase transition
is outlined.

\begin{acknowledgments}
This paper is dedicated to the memory of Vitaly Ginzburg (1916-2009), 
a man of strong physical intuition
whose 
research methods and
style of thinking still remain in-demand nowadays.
I acknowledge thoughtful discussions with Merab Gogberashvili
and other participants
of the Memorial Gamow'105 conference in Odessa
as well as the comments by Alexander Avdeenkov and Dmitry Churochkin
on the properties
of Bose-Einstein condensates.
The instructive debates with Frederik Scholtz
about the problem of choice of proper physical degrees of freedom 
in quantum theory
are acknowledged too.
The correspondence from Vladimir Dzhunushaliev was very helpful as well.
This work was supported under a grant of the National Research Foundation of South Africa. 
\end{acknowledgments}

\appendix*
\scn{Logarithmic \schrod equation}{sec-app}

There exist at least two ways of how 
the logarithmic \schrod equation (LogSE) can be
introduced.
The chronologically first one is based on the separability 
argument - the LogSE is the only local \schrod equation 
(apart from the conventional linear one)
which preserves the separability
of the product states:
the solution of the LogSE for a composite system is a product
of the solutions for uncorrelated subsystems \cite{BialynickiBirula:1976zp}.
The second way is based on the arguments closely related
to open quantum systems and quantum information 
theory \cite{brash91} which is relatively
less known and thus deserves to be reminded here.

Consider a multi-particle (sub)system 
whose dynamics is described by the Hamiltonian-type
operator $\hat{{\bf H}}$.
Besides, this subsystem is in a contact
with its environment such that there is an exchange 
of energy and information.
The state of the system is described by the
vector $|\Psi\rangle$.
If the Hamiltonian 
does not depend on wave function then 
in the \schrod coordinate representation
we recover the linear differential equation for $\Psi$.

However, in general the interactions between the particles
comprising the subsystem depend on the distribution $|\Psi|^2$
of the particles in the configuration space.
To determine this distribution, i.e., to extract, transfer and store the information
in a particular configuration of matter,
one requires certain amount of energy per bit, call it
$\varepsilon $.
The information acquired upon measurement of the state
is proportional to the logarithm of the probability
of an outcome $\Psi$, i.e., 
\be
I_\Psi = -\log_2 (\Omega |\Psi|^2) = -  \ln (\Omega |\Psi|^2)/ \ln 2,
\ee
and the associated entropy of the subsystem is given by
\[
S_\Psi = - k_B \langle \Psi | \ln{(\Omega |\Psi|^2)}|\Psi \rangle 
,
\]
where $k_B$ is the Boltzmann constant.
This entropy minimizes on delta-like distributions and maximizes on uniform
ones.
Here the normalization factor $\Omega $ defines a measurement
reference for the entropy because for continuous systems 
the latter is not absolute.
For instance, one could establish the reference entropy as that for a uniform
distribution hence
if the subsystem has fixed volume and the states are box-normalized
then $\Omega$ equals to this volume.

The above-mentioned energy thus brings the contribution
to the Hamiltonian of the form
\be
\hat{{\bf H}}
\to
\hat{{\bf H}}' = \hat{{\bf H}}
-
\varepsilon 
\log_2 (\Omega |\Psi|^2)
,
\ee
and 
the effective temperature which can be
formally associated with this kind of entropy
is given by 
$
T_\Psi
\equiv
(k_B \beta)^{-1} = (\partial E' / \partial S_\Psi )_\Omega
=
\varepsilon / ( k_B \ln 2 )
,
$
where 
$E' = \langle \Psi | \hat{{\bf H}}' | \Psi \rangle $
is the total energy of the system.
Rewriting $\varepsilon $ in terms of $\beta$, we recover
LogSE in  our notations (\ref{e-xmain}).
For stationary states one can write it in the form
\be
\left[
\hat{{\bf H}}
-
\beta^{-1} \ln (\Omega |\Psi|^2)
\right]
 \Psi  = E'  \Psi 
,
\ee
whereas  the free energy is given by $E =  
\langle \Psi | \hat{{\bf H}} | \Psi \rangle = E' - T_\Psi S_\Psi $.
Unlike the free energy, the energy $T_\Psi S_\Psi$
is engaged in handling the information $I_\Psi$
and thus unavailable to do dynamical work.

The \schrod equations of such type are suitable for describing
subsystems in which the information is not conserved
but being exchanged with environment.
Therefore, they can not be naively applied to systems
without any kind of irreversibility
hence the negative results of 
the experiments \cite{logexp} are not surprising.
On the other hand,
in a theory of
quantum gravity 
this question is still far from being settled  \cite{Preskill:1992tc}.
Besides, one can notice that the logarithmic term describing the information
exchange between a system and its environment 
plays the role
similar to that of the chemical potential in condensed matter systems.
This fulfills the condition for
the condensed-matter-type approach being eligible for description
of the physical vacuum \cite{Volovik:2006bh}.

To conclude, we write down the most important properties of LogSE:
\bit
\item Separability of noninteracting subsystems (as in the linear theory):
the solution of the LogSE for the composite system is a product
of the solutions for the uncorrelated subsystems;
\item Energy is additive for noninteracting subsystems (as in the linear theory);
\item Planck relation holds as in the linear theory;
\item All symmetry properties of the many-body wave-functions with respect to 
permutations of the coordinates of identical particles are preserved in time,
as in the linear theory;
\item Superposition principle is relaxed to the weak one: the sum of solutions
with negligible overlap is also a solution;
\item Free-particle solutions, called \textit{gaussons}, have the coherent-states form,
and upon the Galilean boost they become the uniformly moving
Gaussian wave packets modulated by the de Broglie plane waves;
\item Expressions for the probability density and current are the same
as in the linear theory.
\eit
All these properties except the last one and, perhaps, second last and third last ones, are unique to LogSE
among all other local nonlinear \schrod equations. 
Besides, many of these features are pertinent to the linear
\schrod equation which makes the logarithmic one a ``minimal'' nonlinear
modification in a sense.

\def\AnP{Ann. Phys.}
\def\APP{Acta Phys. Polon.}
\def\CJP{Czech. J. Phys.}
\def\CMPh{Commun. Math. Phys.}
\def\CQG {Class. Quantum Grav.}
\def\EPL  {Europhys. Lett.}
\def\IJMP  {Int. J. Mod. Phys.}
\def\JMP{J. Math. Phys.}
\def\JPh{J. Phys.}
\def\FP{Fortschr. Phys.}
\def\GRG {Gen. Relativ. Gravit.}
\def\GC {Gravit. Cosmol.}
\def\LMPh {Lett. Math. Phys.}
\def\MPL  {Mod. Phys. Lett.}
\def\Nat {Nature}
\def\NCim {Nuovo Cimento}
\def\NPh  {Nucl. Phys.}
\def\PhE  {Phys.Essays}
\def\PhL  {Phys. Lett.}
\def\PhR  {Phys. Rev.}
\def\PhRL {Phys. Rev. Lett.}
\def\PhRp {Phys. Rept.}
\def\RMP  {Rev. Mod. Phys.}
\def\TMF {Teor. Mat. Fiz.}
\def\prp {report}
\def\Prp {Report}

\def\jn#1#2#3#4#5{{#1}{#2} {\bf #3}, {#4} {(#5)}} 

\def\boo#1#2#3#4#5{{\it #1} ({#2}, {#3}, {#4}){#5}}



\end{document}